\documentclass[12pt]{article}
\usepackage{pdproc} 
\usepackage{graphicx}

\begin{document}

\renewcommand{\thefootnote}{\alph{footnote}}
  
\title{DIRECT DARK MATTER SEARCHES:\\
FITS TO WIMP CANDIDATES\footnote{Invited review talk at the XIV International Workshop on ``Neutrino Telescopes", March 15 to 18, 2011,  Venice, Italy}}

\author{GRACIELA  B. GELMINI}

\address{Department of Physics and Astronomy,  University of California Los Angeles\\
405 Portola Plaza, 
 Los Angeles, CA 90095, USA.\\
 {\rm E-mail: gelmini@physics.ucla.edu}}


\abstract{After a brief introduction to dark matter in general and to WIMPs as candidates, we review recent results of direct dark matter searches. We concentrate on older and more recent hints pointing to light  WIMP's with mass below 10 GeV.}
   
\normalsize\baselineskip=15pt

\section{Dark Matter: what we know}
  We know a lot about dark matter (DM). We know its abundance  in the Universe to a few percent level, we know  most of it is not baryonic and we know that it is not explained  within the Standard Model (SM) of elementary particles. WMAP 7 year Cosmic Microwave Background anisotropy data
   combined with Type Ia supernovae and the Baryon Acoustic Oscillations data, give a total matter density fraction $\Omega_m= 27.8 \pm 1.5\%$ and, in agreements with Big-Bang Nucleosynthesis (BBN) data,  a baryon abundance which is only 4-5\% of the total density, $\Omega_{\rm b}=4.61 \pm 0.15\%$, so the rest of the matter is in DM,
  $\Omega_{\rm DM} = 23.2 \pm 1.3\%$~\cite{Komatsu:2010fb}. We know also from structure formation arguments that  the bulk of the DM can only be either Cold (CDM) or Warm (WDM) (i.e. non-relativistic or becoming non-relativistic when the temperature of the Universe was  $T\simeq$ keV).  Among the particles of the SM only neutrinos are part of the DM and they are Hot DM (i.e. relativistic when  $T\simeq$ keV). There are no WDM or CDM candidates in the SM, but there are many in all SM extensions.

Because of spontaneous symmetry breaking arguments totally independent of the DM issue, we do expect new physics beyond the SM to appear at the electroweak scale  explored by the Large Hadron Collider (LHC) at CERN. Physics beyond the SM is required by the DM and expected at the electroweak scale, but both new physics may or may not be related. Thus the experiments at the LHC and the searches for the DM in our galactic halo are independent and complementary.

 We will concentrate in the following on Weakly Interacting Massive Particles (WIMPs) as DM candidates. 

\section{WIMPs: earliest relics}
  The argument showing that WIMPs  are good DM candidates is old and well known. The density per comoving volume of non relativistic  particles in equilibrium in the early Universe decreases exponentially with decreasing temperature, due to the Bolzmann factor, until the reactions which change the particle number become ineffective.  
 At this point, when  the annihilation rate  becomes smaller than the Hubble expansion rate, 
  the WIMP number per comoving volume becomes constant. This  is the  moment of chemical decoupling or freeze-out. For larger annihilation cross sections $\sigma_{\rm annih}$, freeze-out happens later, when WIMP densities are smaller. If there is no subsequent change of entropy in matter plus radiation, the present relic density is $\Omega_{\rm standard} h^2 \simeq 0.2 \left[ 3 \times 10^{-26} {\rm cm}^3/{\rm s} \right] \left< \sigma_{\rm annih} v \right>^{-1} $, which for weak cross sections  $\sigma_{\rm annih}= 3 \times 10^{-26} {\rm cm}^3/{\rm s}$ gives the right order of magnitude of the DM density $\Omega \simeq 0.2$ and a freeze-out temperature  $T_{\rm f.o.} \simeq m/20$  for a WIMP of mass $m$.

  BBN is the earliest episode (200 s  after the Bang) from which we have a trace, the abundance of light elements D, $^4$He and $^7$Li.   It is enough that the highest temperature of the radiation dominated period in which BBN happens is larger than 4 MeV~\cite{hannestad} for BBN and all the subsequent  history of the Universe to proceed as usual. Notice that  $T_{\rm f.o.}>4$ MeV  for any WIMP with $m> 80$ MeV.
Thus, if WIMPs exists, they are remnants from before BBN,  an epoch from which we have no data.

To compute the WIMP relic density we must make assumptions about the pre-BBN epoch. The standard assumptions are reasonable, but are just that, assumptions. We assume that the Universe was radiation dominated at decoupling, that the entropy of matter and radiation is conserved, that WIMPs are produced thermally (i.e. via interactions with the particles in the plasma)  and were in 
equilibrium before they decoupled  (usually also
 that there were no particle-antiparticle WIMP asymmetries). These are just assumptions, which do not hold in many cosmological models and in those the WIMP relic density can be very different than what our standard computations indicate (see e.g. Ref.~3). So, no matter the properties of a DM particle candidate that we may uncover in accelerators, we will not be sure that it constitutes the DM unless we find it in the dark halos of galaxies! In fact, if ever discovered, WIMPs 
would  for the first time give  information on the pre-BBN epoch of the Universe. Thus, WIMP DM searches are complementary to LHC searches.
  
\section{WIMP searches}
  
  Indirect DM searches look for WIMP annihilation (or decay) products,  e.g. neutrinos from the center of the Sun or the Earth  or anomalous cosmic rays, such as e$^+$ and $\bar{p}$,   or $\gamma$-rays from the galactic halo  or the galactic center.    Direct DM searches look for energy deposited within a detector by the collisions  of  WIMPs belonging to the dark halo of our galaxy. There have been many DM ``hints" in both types of searches. I will concentrate on direct searches and mention recent data of (in alphabetic order)   CDMS, CoGeNT,  CRESST II,  DAMA,  XENON 10, XENON 100.  
  
 The  momentum exchange $q$ in each collision would be small enough that WIMPs would interact coherently with a nucleus within a detector. The differential event rate in direct DM detectors depends on 
\begin{equation}  
 \frac{dR}{dE} =
\frac{{\rho} {\sigma(q)}}{2 m \mu^2} 
\int_{v>v_{\rm min}}
\frac{f({\bf v},t)}{v} d^3v, 
\end{equation}  
where $\mu= mM/ (m+M)$ is the WIMP-nucleus reduced mass. The minimum WIMP speed necessary to impart a recoil energy $E$ is  $v_{\rm min}=\sqrt{ME/2\mu^2}$ for elastic scattering.
   
    The rate depends on particle properties through the mass $m$ and the scattering cross section $\sigma(q)$.  For WIMPs with spin-independent (SI) interactions,  the cross section has an enhancement of a factor proportional to the atomic number squared $A^2$, i.e. $\sigma(q) = \sigma_0 F^2(q)$ where $\sigma_0 = \bigl[ \langle Z f_p+ (A-Z) f_n \bigr]^2 (\mu^2/\mu_{\rm p}^2)  \sigma_{\rm p}$ and the form factor $F^2(q)$ takes care of the departure from coherence. If the WIMP couplings to protons and neutrons are the same $f_p=f_ n$, then  $\sigma_0 = A^2 (\mu^2/\mu_{\rm p}^2) \sigma_{\rm p}$.  In this case only two parameters remain, the WIMP mass $m$  and   the WIMP-proton cross section $\sigma_{\rm p}$ ($\mu_{\rm p}$ is the WIMP-proton reduced mass).       For  spin-dependent (SD) interactions $\sigma(q)$ depends on the total spin of the nucleus and is typically a factor $A^2$ smaller.  I will concentrate on SI interactions, where most of the action has been in recent months.

  The rate depends on astrophysical parameters through the local DM density $\rho$ and
   velocity distribution $f({\bf v},t)$. These depend on the model of the dark halo of our galaxy. In order to compare the results of different experiments usually the Standard Halo Model (SHM) is used, in which $\rho=0.3$ GeV/cm$^3$ and $f ({\bf v},t)$ is a  truncated Maxwellian distribution with zero average velocity and escape speed $v_{esc}\simeq$ 500 - 650 km/s  with respect to the Galaxy. Thus, the average velocity of WIMPs with respect to Earth is mostly due to the velocity of the Sun around the Galaxy $v_\odot\simeq$220 km/s.
We expect the actual halo to deviate somewhat from this simplistic model. The local density and velocity could be actually very different if the Earth is within a DM clump or stream, which is unlikely~\cite{Vogelsberger},  or if there is a ``Dark Disk"~\cite{Read} in our galaxy. 
 
  WIMP interactions in crystals produce mostly phonons. Only a fraction $Q$ (the ``quenching factor") of the recoil energy goes into ionization or scintillation. E.g. $Q_{\rm Ge} \simeq 0.3$, $Q_{\rm Si} \simeq 0.25$, $Q_{\rm Na} \simeq 0.3$, $Q_{\rm I} \simeq 0.09$. In noble gases, a similar factor,  $L_{\rm eff}$, measures the scintillation efficiency of a WIMP relative to a $\gamma$. There are large experimental uncertainties in the  determination of  these parameters at low energies.

 There are many direct DM detection experiments running or in construction  or in the stage of research and development. They use different target materials and different detection strategies. Single channel techniques measure only one of the effects produced by the recoiling nucleus  hit by a DM WIMP, usually phonons (or heat), ionization or scintillation. Among the experiments measuring only ionization (in Ge, Si or CdTe) are IGEX, HDMS, GENIUS, TEXONO and CoGeNT, among those using only scintillation (in NaI, Xe, Ar, Ne or CsI) are DAMA, NAIAD, DEAP/CLEAN, XMASS and  KIMS and among those using phonons  (in Ge, Si, Al$_2$O$_3$ or TeO$_2$) are CRESST-I, Cuoricino and CUORE. Threshold detectors search instead for bubbles produced by the energy deposited in a WIMP collision, either in a superheated bubble chamber, as done by COUPP, or  with superheated freon, C$_4$F$_{10}$, droplets suspended in a CsCl gel, as in PICASSO. Several experiments use hybrid detector techniques in which the relative intensity of two different effects is used to discriminate between nuclear recoils and the background. Experiments such as CDM, SuperCDMS, EDELWEISS and EURECA use ionization and phonons (in Ge or Si); ZEPLIN, XENON 10, XENON 100, LUX, WARP, ArDM and DarkSide  use ionization and scintillation (in Xe, Ar or Ne); CRESST-II uses scintillation and photons
 (in CaWO$_4$). Some of these experiments, such as CDMS or EDEWEISS use also timing of the signal to provide further discrimination. Liquid noble-gas detectors, using Xe, Ar or Ne, either in a single (liquid) phase and measuring scintillation, or in double (liquid-gas) phase and measuring ionization and scintillation are very promising, because their mass can in principle be easily upgraded to several tonnes. Directional WIMP dark matter detectors, for which there are several prototypes,  are also extremely interesting.

\section{Light WIMPs: early hints and bounds}
 
 Let us start with the oldest ``hint".  The DAMA
collaboration has found an annual modulation in their data compatible
with the signal expected from DM particles bound to our
galactic halo and the SHM, due to the motion of the Earth around the Sun. The  annual modulation seen by the DAMA/NaI experiment~\cite{Bernabei:2003za} was confirmed by 
DAMA/LIBRA in 2008~\cite{Bernabei:2008yi} and in 2010~\cite{Bernabei:2010mq}.
The 7 years of data of DAMA/NaI showed a 6$\sigma$ modulation signal~\cite{Bernabei:2003za}.  Now, the combined 13 years of data  (with a very impressive exposure,  1.17 ton$\times$year) show an 8.9$\sigma$ modulation signal~\cite{Bernabei:2010mq}.

Are the DAMA results compatible with those of all negative searches? There are many aspects to this question and  I will concentrate on WIMPs which scatter elastically through SI interactions with nuclei and the SHM. 

\begin{figure*}[t]
\centering
\includegraphics[width=0.35\textwidth]{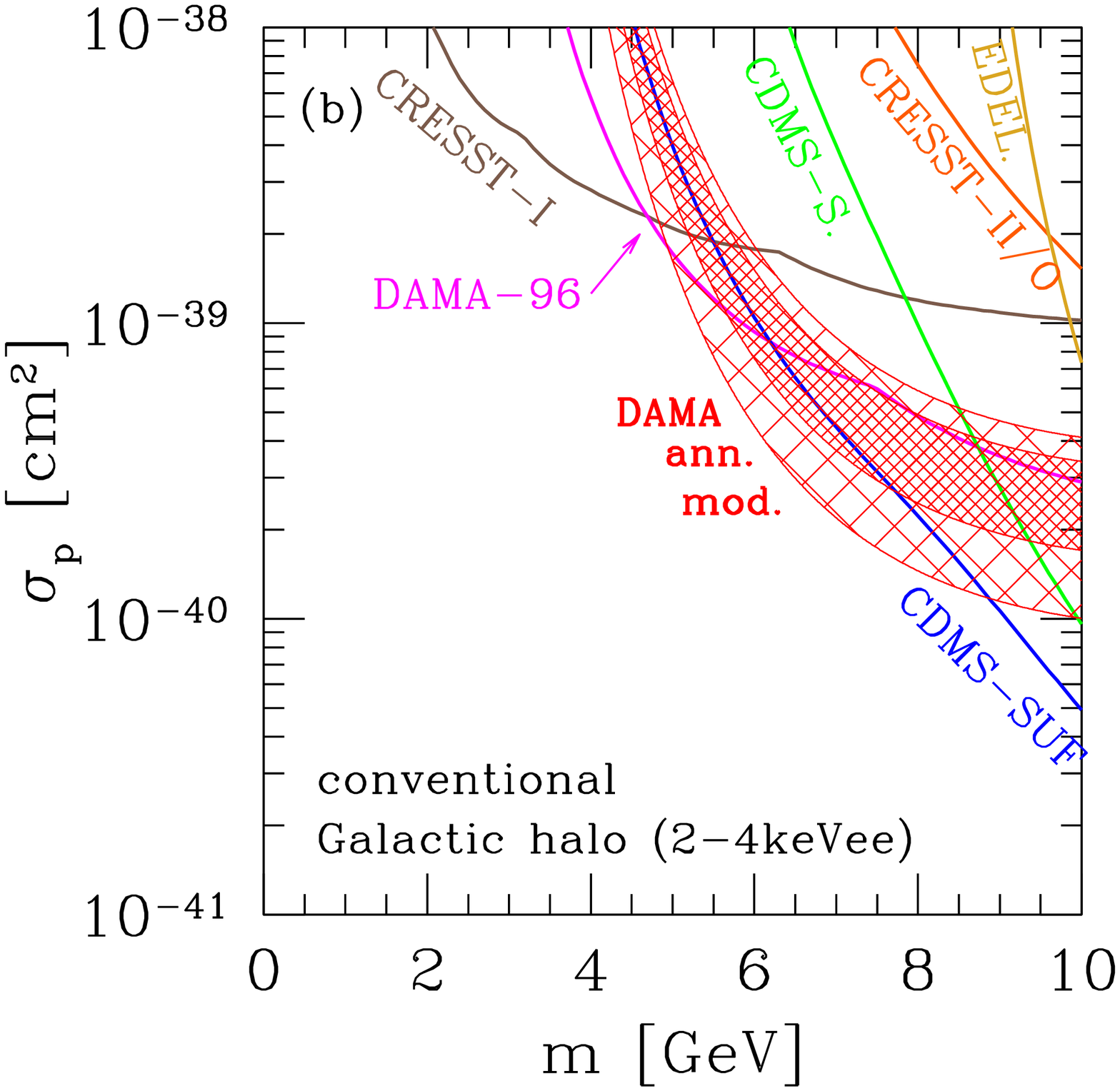}$~~~~~~~~~~~~$
\includegraphics[width=0.33\textwidth, angle=90]{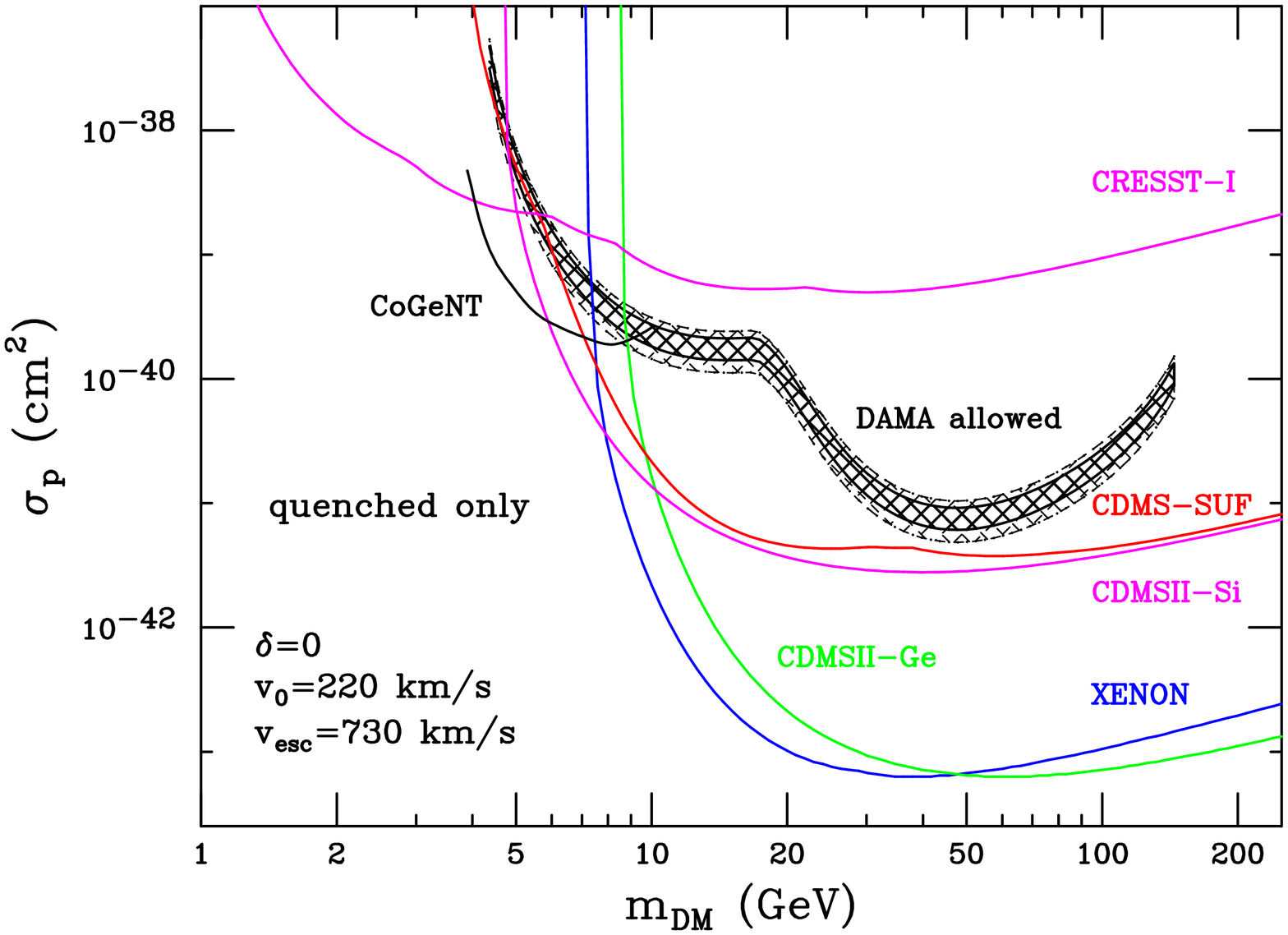}
\vspace{-10pt}
\caption{Light WIMPs region compatible with 1.a (left)  the DAMA data in 2005~\protect\cite{Gelmini:2004gm},   1.b (right) the DAMA/LIBRA data in 2008~\protect\cite{Petriello:2008jj}, both  obtained with  a ``raster scan" in mass as in Ref. 9.
}
\vspace{-10pt}
 \label{F1}
\end{figure*}

\begin{figure*}[t]
\centering
\hspace{-0.5cm}
\includegraphics[width=0.26\textwidth]{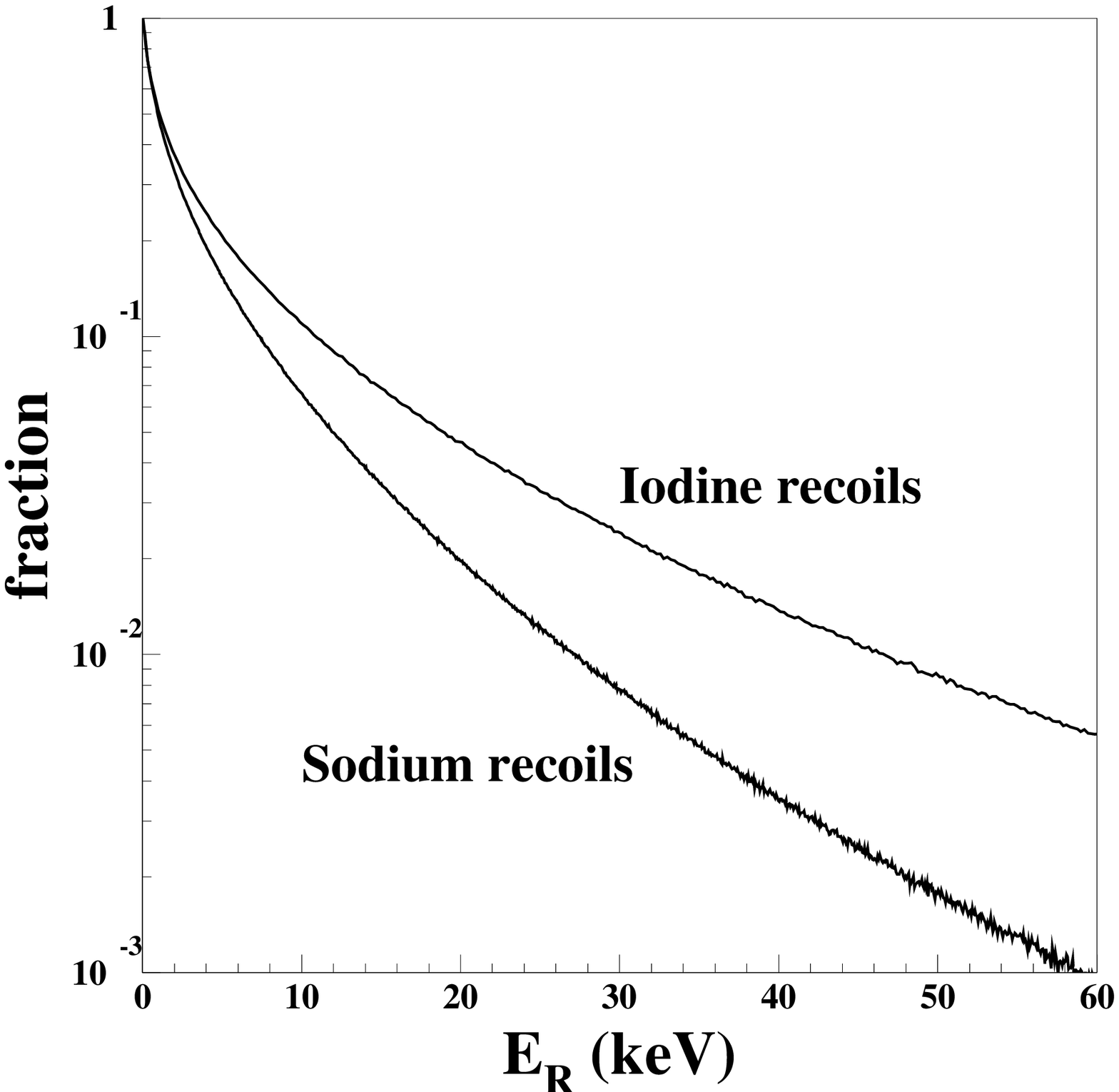}
\includegraphics[width=0.49\textwidth]{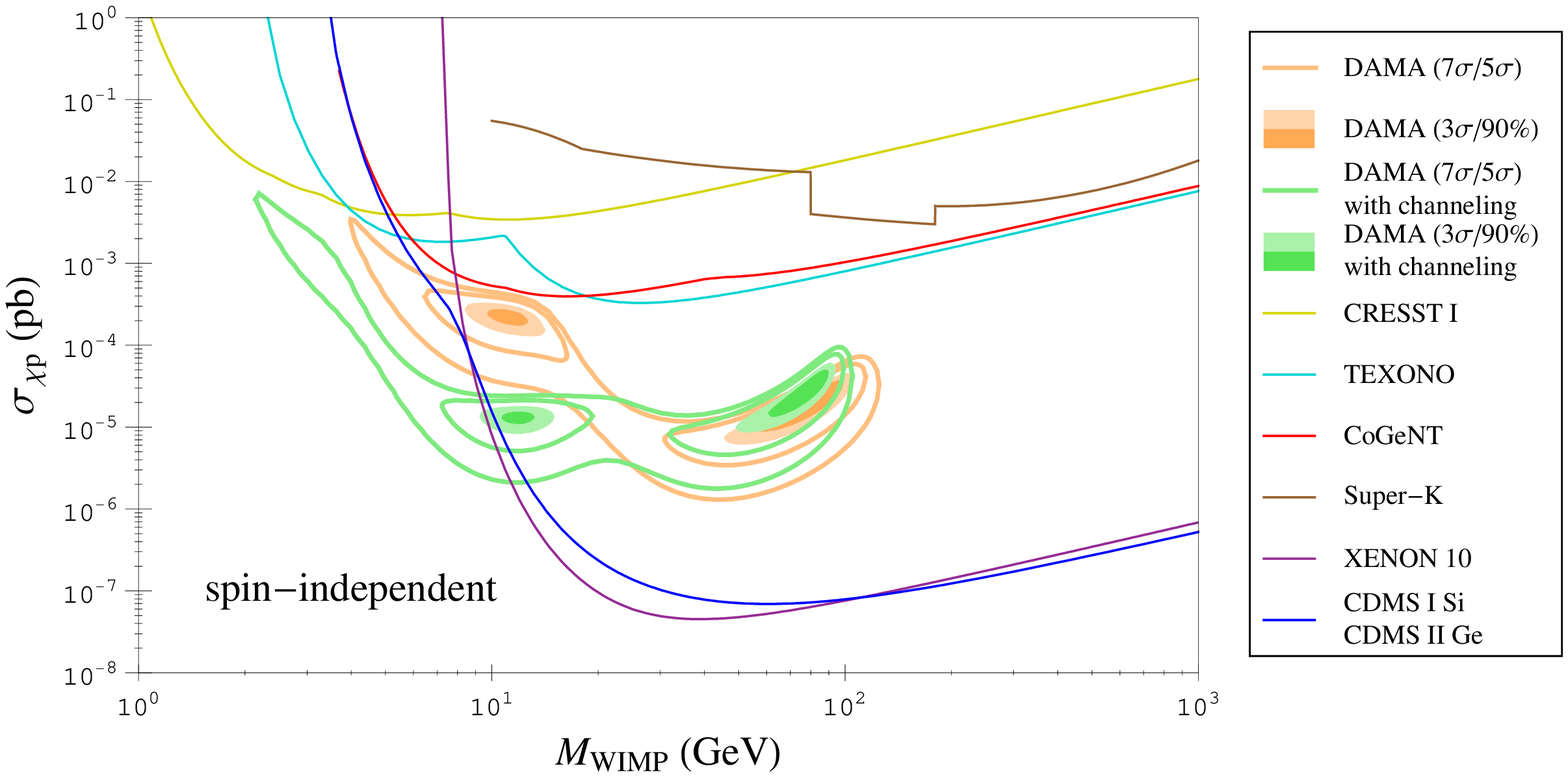}
\vspace{-10pt}
\caption{\ref{F2}.a (left) Channeling fractions of Ref.~15 (in which blocking effects were not included) and \ref{F2}.b (right) fits to the DAMA data~\protect\cite{Savage:2008er} including (green region) and not including (orange region) these channeling fractions (higher and lower  $\sigma_p$ regions for $m \sim 10$ GeV due to Na and  I respectively).}
 \label{F2}
\end{figure*}

Due to a theoretical prejudice the early DAMA analysis produced a  region of compatible WIMPs only for masses above 30 GeV, which was excluded in 2004 by CDMS-Soudan and Edelweiss. However,  at TAUP2005 in Spain I used the spanish aphorism ``Los muertos que vos matais gozan the buena salud" (``The dead  you kill are in good health")  in a talk whose main point was  that, even with the usual SHM for the dark halo, the DAMA claim had not yet been rejected by other experimental results for light WIMPs~\cite{Gelmini:2004gm}. WIMPs in a standard halo in the mass range 5 to 9 GeV  and with WIMP-proton scattering cross section $\sigma_p$ in the range $10^{-40}$ to $10^{-39}$~cm$^2$ as shown in Fig.~\ref{F1}.a (a corner of the region already found by DAMA as a superposition of many halo models- see Fig.~28 of Ref.~6)
 provided a good fit to the DAMA data and were above the speed threshold for interaction with Na in  DAMA and below the WIMP speed threshold for CDMS and EDELWEISS (simply because Na is lighter than Ge)~\cite{Gelmini:2004gm}  so that the positive and negative detection results could be compatible  (see Fig.~\ref{F1}.a).  Models with light neutralinos as WIMPs with mass as low as 6 GeV already existed  at the time~\cite{bottino} (although with cross sections about one order of magnitude smaller), but Ref.~9 proceeded in a pure phenomenological manner (and many light WIMP particle models have been proposed now; see e.g. Refs. 11,12  and references therein).
Subsequently, the CDMS collaboration made an effort to lower its threshold.  They used a small exposure of Si, 12 kg d, with a low threshold (7 keV) to produce in 2005 what remained until recently one the best bounds on light WIMP's~\cite{CDMS-06} (shown in Figs.~\ref{F1}.b, \ref{F2}.b and \ref{F4}.a as CDMS Si).

The DAMA/NaI data were given in only two independent bins (2-4 keVee or 2-6 keVee and 6-14 keVee). Thus in Ref. 9  a one parameter fit was performed to obtain the range of $\sigma_p$ for each value of  WIMP mass $m$ (in what is sometimes called a ``raster scan" in $m$). Shortly after the first DAMA-LIBRA results were announced in 2008, the same analysis of Ref.~9 was repeated with the new data in Ref.~14, which concluded that the region of light WIMPs was rejected (see Fig.~\ref{F1}.b) unless channeling as evaluated by the DAMA collaboration in 2007~\cite{Bernabei:2007hw} was included (in which case the region of light WIMPs still remaining was due to channeled I recoils).  The annual modulation data of DAMA/LIBRA was given not only in two bins but in 36, thus a raster scan was not justified and using all data bins the conclusions on light WIMPs change somewhat (see e.g.  Fig~\ref{F2}.b~\cite{Savage:2008er}).  The spectral modulation amplitude information favors heavier WIMPs (because the modulation amplitude in the lowest energy bin is smaller than in the subsequent bins). The most preferred DAMA regions of any WIMP mass are ruled out to 3$\sigma$. However, for WIMP masses under 10~GeV  some parameters outside these regions still yield a reasonable fit to
the DAMA data and as of 2008 were  compatible with all 90\% C.L. upper limits from negative searches  (see e.g.  Fig~\ref{F2}.b~\cite{Savage:2008er}).
Understanding channeling in direct DM detectors  was very important  to  ascertain DAMA's compatibility with other data sets, particularly at low masses. Including channeling as evaluated by the DAMA collaboration (the channeling fraction shown in Fig~\ref{F2}.a~\cite{Bernabei:2007hw}) shifted the region of compatible light WIMPs to lower values of $\sigma_p$ by more than one order of magnitude (as shown in 
Fig~\ref{F2}.b~\cite{Savage:2008er}).
 \begin{figure*}[t]
\centering
\hspace{-0.5cm}
\includegraphics[width=0.31\textwidth]{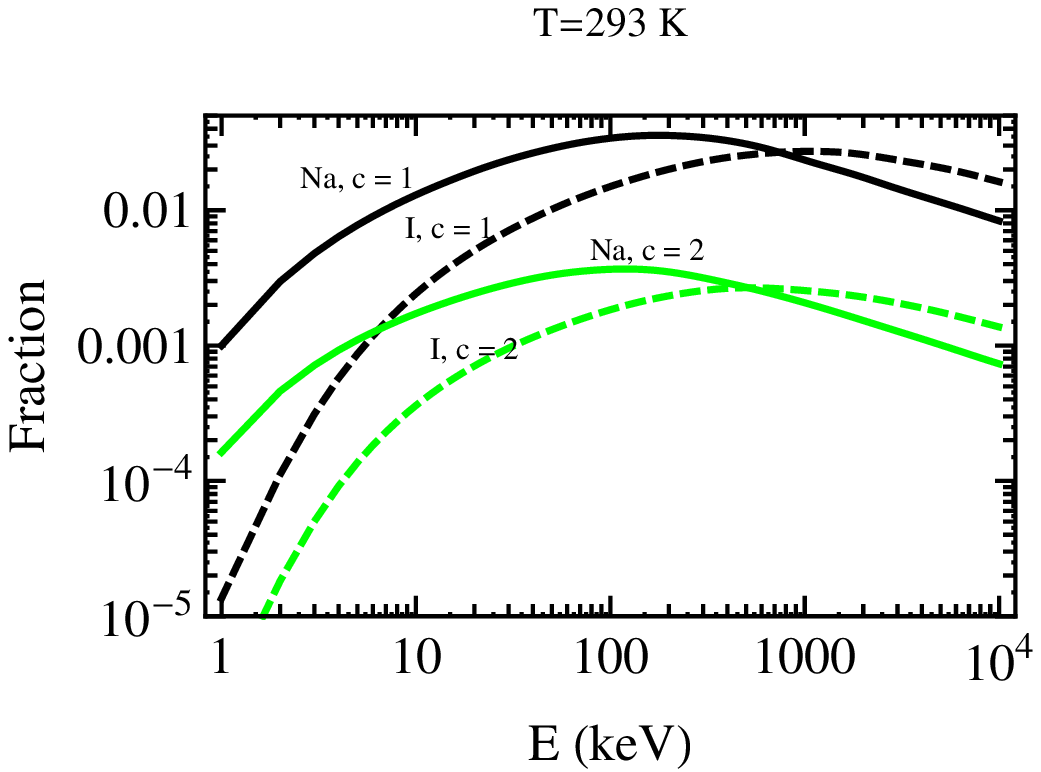}
\includegraphics[width=0.49\textwidth]{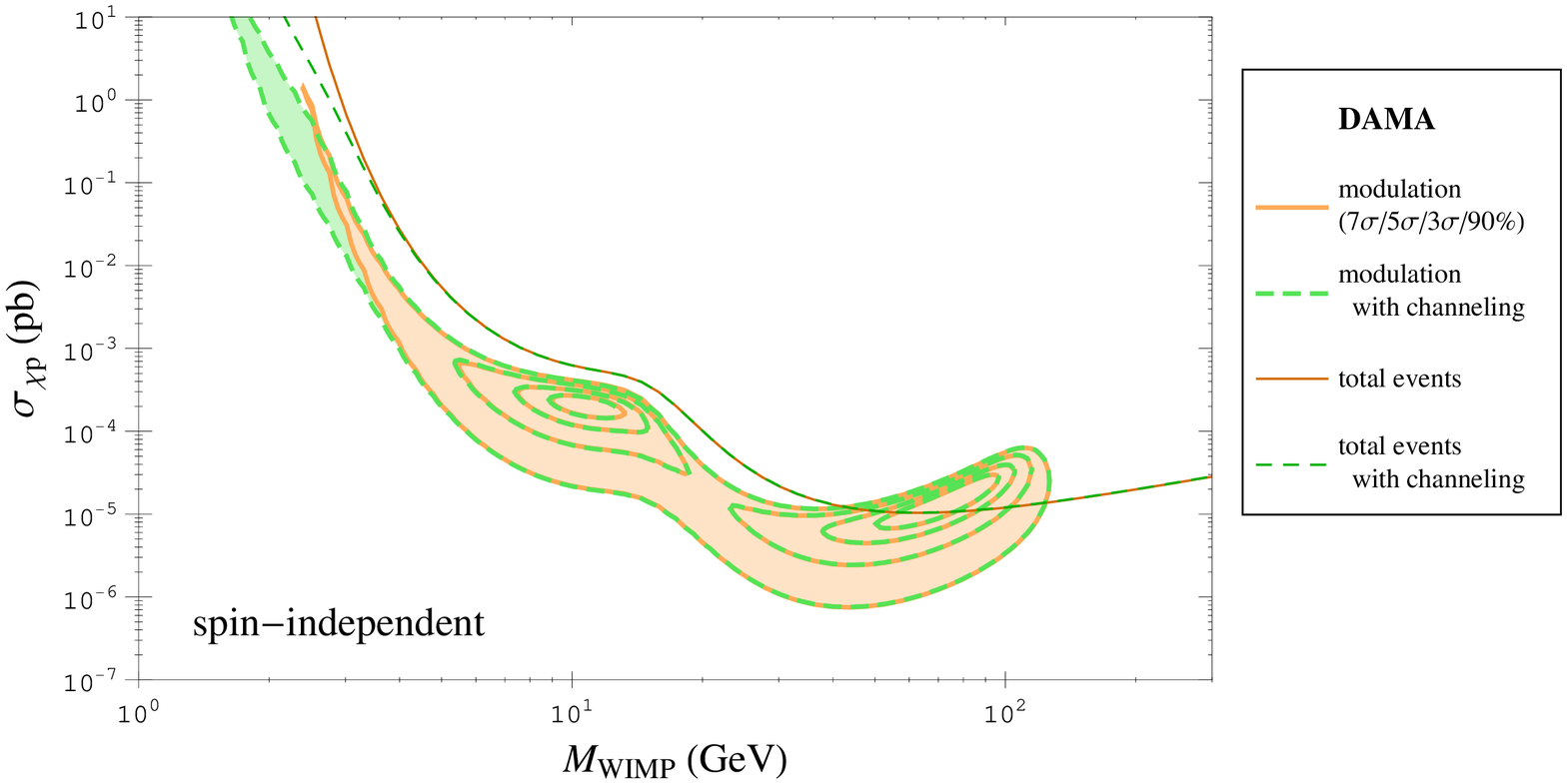}
\vspace{-10pt}
\caption{\ref{F2-PRIME}.a (left) Upper bounds to the channeling fractions as evaluated in Ref. 21 and  \ref{F2-PRIME}.b (right) fits to the DAMA annual modulation data~\protect\cite{Savage:2010tg} including (green region) and not including (orange region) these channeling fractions (with $c=1$) showing no difference at less than 7$\sigma$.}
\vspace{-10pt}
 \label{F2-PRIME}
\end{figure*}

Channeling and blocking effects in crystals refer to the orientation dependence of the penetration of ions in crystals.  Channeling happens when ions moving  in the crystal along symmetry axes and planes
suffer a series of small-angle scatterings  that maintain them in the open ``channels". Channeled ions do not get close to a lattice site, where they would suffer a  large-angle scattering.  Blocking consists in a reduction of the flux of ions originating in lattice sites along symmetry axes and planes of a crystal. Both effects were discovered in the 1960's~\cite{Gemmell:1974ub} and are extensively  used in crystallography,  in measurements of short nuclear lifetimes, in the production of polarized beams etc. Channeling must be avoided in the implantation of   B, P and As atoms  in Si crystals to make circuits~\cite{Hobler}, one of the few applications at low energies, below MeV, the energies of interest here.   The channeling effect in NaI(Tl) crystals was first observed by Altman et al.~\cite{altman} in 1973 who observed that channeled ions produce more scintillation light because they loose practically all their energy via electronic stopping rather than nuclear stopping. 
 When ions recoiling after a collision with a WIMP are channeled, they give all their energy to electrons, so for them the quenching factor is  $Q \simeq1$. 
 
 The  potential importance of channeling for direct  DM detection was first pointed out  by 
 Drobyshevski~\cite{Drobyshevski:2007zj}  in 2007 and soon after by the DAMA collaboration~\cite{Bernabei:2007hw}. The DAMA paper~\cite{Bernabei:2007hw}  gave an estimate of the channeling fraction of recoiling ions  in which the fraction  grew with decreasing energy to be $\simeq 1$ close to 1 keV (Fig.~\ref{F2}.a). However this estimate  did not take into account blocking~\cite{BGG}.   When colliding with WIMPs, ions are ejected from lattice sites, thus blocking is important and channeling fractions are reduced. In fact in a perfect lattice no recoiling ion would be channeled (as no channeled ion gets close to a lattice site, what Lindhard called the ``rule of reversibility", based on time reversal of paths), but due to lattice vibrations the collision with a WIMP may happen while  an atom is displaced with respect to the row or plane where it belongs. If it is initially far enough it may be channeled.
   In Ref.~21  analytic models developed in the 1960's and 1970's were used to study channeling in direct DM detectors and upper bounds to  channeling fractions  were obtained. The effect is strongly temperature dependent, and a parameter $c$ (in Fig.~\ref{F2-PRIME}.a), a number between 1 and 2 enters in the evaluations of the temperature dependence. Using these fractions (with $c=1$), shown in
   Fig.~\ref{F2-PRIME}.a for NaI at room temperature, 
    the WIMP regions corresponding to the DAMA annual modulation signal with channeling included and not included differ only at 7$\sigma$ (see Fig.~\ref{F2-PRIME}.b~\cite{Savage:2010tg}).

\section{Light WIMPs: recent  hints and bounds}

Several recent potential DM hints (or just background events) were found in different experiments.  The CDMS collaboration found two highly publicized events in Dec. 2009~\cite{Ahmed:2009zw}, most probably background events.  They found 2 events with 0.9 of background expected in 612 kg d of data (a previous similar run had found 0 events with 0.6 of background expected in about 400 kg d of data, which would make 2 events when 1.5 were expected).

\begin{figure*}[t]
\centering
\includegraphics[width=0.59\textwidth]{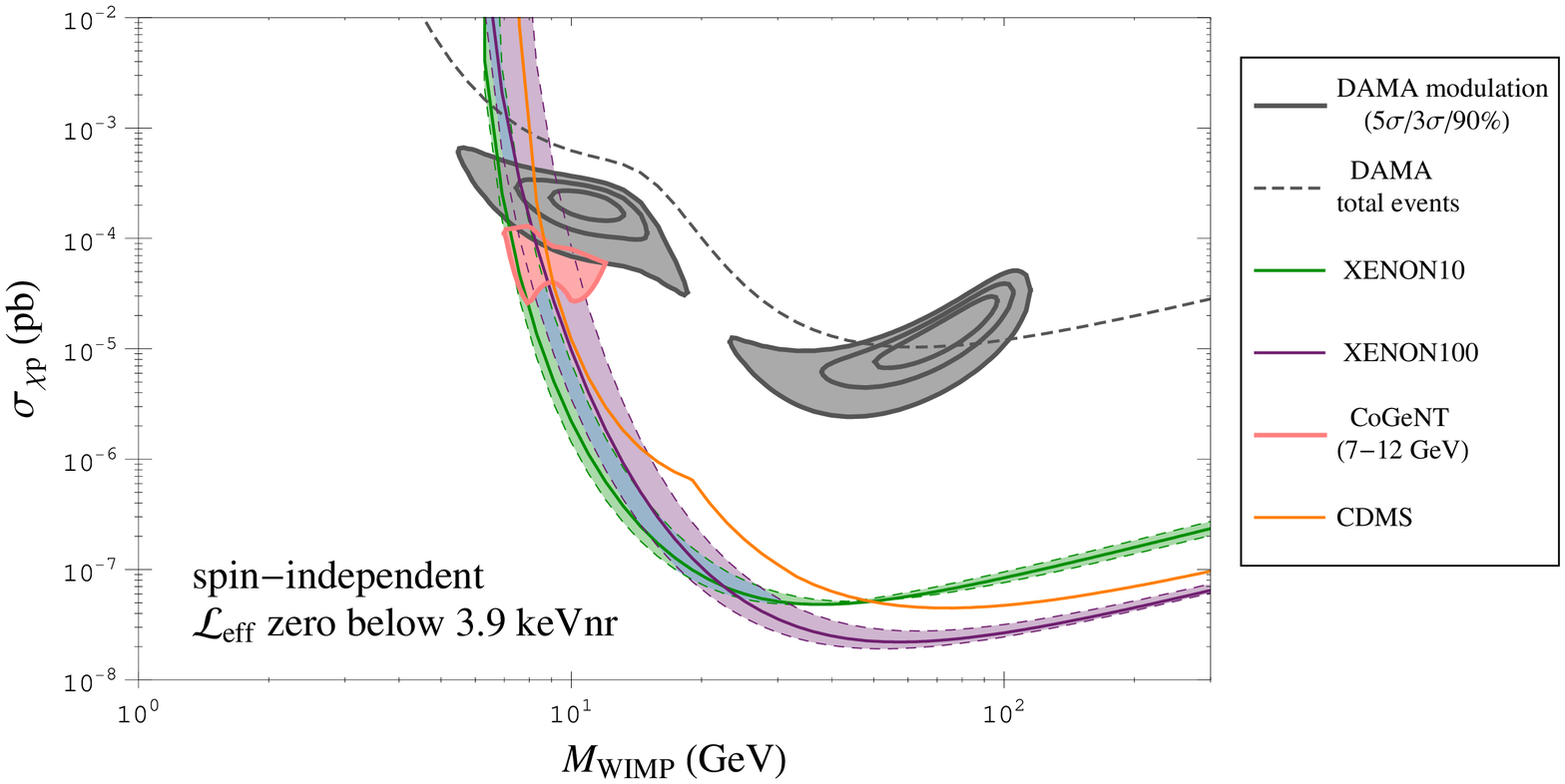}
\includegraphics[width=0.40\textwidth]{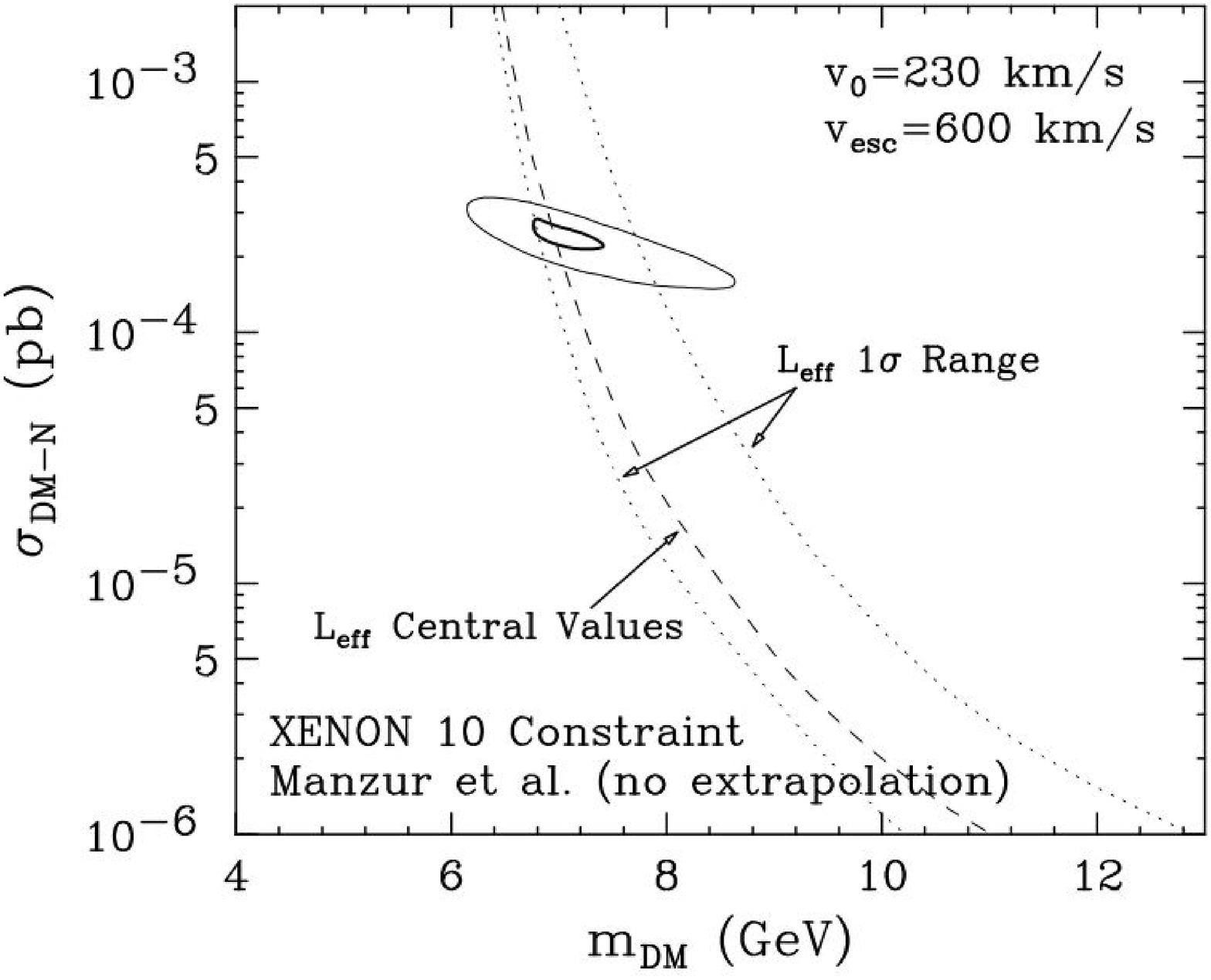}
\vspace{-10pt}
\caption{XENON10 and XENON100 (11 days of data) bounds~\protect\cite{Savage:2010tg}  derived with the extreme assumption that $L_{\rm eff}$=0 below the lowest point measured by Manzur et al.~\protect\cite{Manzur:2009hp} shown in both panels, with \ref{F3}.a-(left) the DAMA (grey) and CoGeNT~\protect\cite{Aalseth:2010vx} (pink) regions~\protect\cite{Savage:2010tg} or \ref{F3}.b-(right) the combined region (assuming $Q_{\rm Na}=0.2-0.4$ and no CoGeNT exponential background)~\protect\cite{Hooper:2010uy}.}
\vspace{-10pt}
\label{F3}
\end{figure*}

In Feb. 2010 the CoGeNT collaboration~\cite{Aalseth:2010vx}, using a small 440g Ge detector in the Soudan Mine with extremely low threshold, 0.4 keVee,  in 56 days of data  found and excess  of irreducible  bulk-like events compatible with the signal of a light WIMP with $m$ close to 9 GeV WIMP and cross section close to 10$^{-40}$ cm$^2$ (pink irregular region~\cite{Aalseth:2010vx}  reproduced in Fig.~\ref{F3}.a). There is an exponential component of the spectrum, the dominant component at energies below 1 keVee, which is compatible with the signal of a WIMP. However, it is not clear if there is an exponential component to the background and if so which is its amplitude. If the amplitude of this potential background component is left as a free parameter in the fit, the best fit of signal plus background is no signal at all. Thus, having a WIMP signal depends on assuming a ``constrained" exponential background, which means that the background is assumed to be a fraction of the exponential rate (like a third, e.g.~\cite{Chang:2010yk}), or required bin-by-bin not exceed the amplitude of the DM signal (e.g.~\cite{Fitzpatrick:2010em}) or assumed to be just zero (e.g.~\cite{Hooper:2010uy}). Both panels of Fig.~\ref{F3} show bounds derived in Ref. 18 from XENON10 and  11 days of data of XENON100 assuming that $L_{eff}$ goes to zero below the lowest energy data point of Manzur et al.~\cite{Manzur:2009hp}, the lowest measurements and thus most conservative choice  until very recently. The bands reflect the 1-$\sigma$ band in these  measurements~\cite{Manzur:2009hp}.  With standard parameter choices the CoGeNT region and Na DAMA regions are close to each other (Fig.~\ref{F3}.a), but they can overlap if  $Q_{\rm Na}=$  0.2 to 0.4  (a reasonable range instead of the usual value 0.3)~\cite{Hooper:2010uy}. The overlap region~\cite{Hooper:2010uy}   is shown Fig.~\ref{F3}.b. The figure shows some small allowed regions (but there are newer data - see below).

\begin{figure*}[t]
\centering
\includegraphics[width=0.37\textwidth]{SCHWETZcdms-all.eps}
\includegraphics[width=0.62\textwidth]{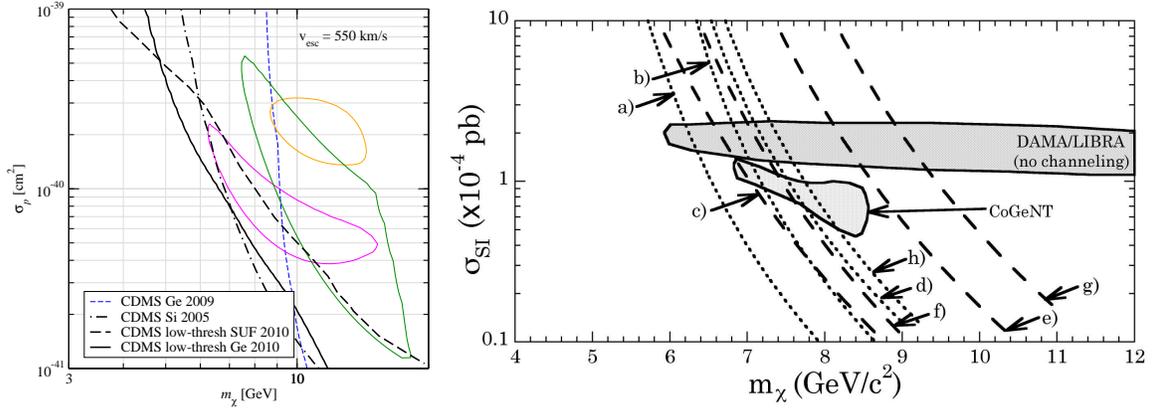}
\vspace{-10pt}
\caption{\ref{F4}.a DAMA Na, CoGeNT 56-day with no exponential background  and potential CRESST II regions (yellow, magenta and green contours respectively) shown together with the 2005 Si CDMS bound~\protect\cite{CDMS-06}  and new low threshold, 2 keV,  bounds from the old CDMS at SUF~\protect\cite{Akerib:2010pv} and CDMS II~\protect\cite{Ahmed:2010wy}. \ref{F4}.b New CoGeNT 15-months region~\protect\cite{Aalseth:2011wp}  (lower $\sigma_p$ region), extended DAMA Na region (with Q$_{\rm Na} =$ 0.2 to 0.4) and different XENON10 and XENON100 bounds  as interpreted by 
J. Collar~\protect\cite{Collar:2011wq}. 
}
\vspace{-10pt}
 \label{F4}
\end{figure*}

Also in Feb. 2010 preliminary results of CRESST II~\cite{Seidel}  from 564 kg d  taken with 9 CaWO$_4$ crystals were announced which show an excess of events in their oxygen recoil band which could be from a neutron background or leakage of $\alpha$ particles or  a light WIMP with mass $m<10$ GeV (for which only recoils in O are above threshold). The 1$\sigma$ DAMA Na region (yellow contour),  CoGeNT 56-day region (magenta contour) and potential CRESST II region (green contour) are all shown together with the old 2005 Si CDMS bound~\cite{CDMS-06}  and new low threshold, 2 keV,  bounds from CDMS (both from the old CDMS at SUF~\cite{Akerib:2010pv} and CDMS II~\cite{Ahmed:2010wy}) are shown in Fig.~\ref{F4}.a~\cite{Schwetz:2010gv}.

In the intervening time between the conference in March 2011 and the production of the written version of this talk in June several important results have appeared. In April 2011,  XENON100  announced new upper bounds based on 100 days of data~\cite{Aprile:2011hi}, obtained using  new measurements of $L_{eff}$~\cite{Plante:2011hw} extending to 3 keVnr, a lower energy than before. New XENON 10 limits appeared shortly  after~\cite{Angle:2011th}, derived using only their ionization signal, S2 (usually the ratio of scintillation over ionization, S1$/$S2,  is used as the main signal but for low mass WIMPs S2 alone allows for a lower threshold, 1.4 keV).   And on May 5, 2011 (at the APS  Spring Meeting in Anaheim) CoGeNT announced a 2.8$\sigma$  annual modulation  signal in the exponential part of the spectrum in 15 months of 
data~\cite{Aalseth:2011wp}. Interpreted as due to a light WIMP produces the lower $\sigma_p$ region shown in Fig.~\ref{F4}.b~\cite{Collar:2011wq} which does not  overlap with the extended DAMA region (with Q$_{\rm Na} =$ 0.2 to 0.4 but would with 0.40-0.45~\cite{Hooper:2011hd}) also shown in Fig.~\ref{F4}.b, together with different XENON10 and XENON100 bounds  as interpreted by Juan Collar~\cite{Collar:2011wq}.

DM searches  are advancing fast and lots of data necessarily  lead to many hints. Hopefully at some point several of them will point to the same DM candidate. If the DAMA  modulation is due to DM, a DM signal should be found by another experiment, as maybe it has in CoGeNT.
Light WIMP's  are among the most promising candidates to make the DAMA annual modulation data compatible with all other negative searches. Necessarily a
light WIMPs signal would be close to threshold where backgrounds are difficult to understand.
In the near future CoGeNT and also CRESST II will eventually understand better their backgrounds and, in the case of CoGeNT its annual modulation. Certainly more experimental as well as theoretical work is needed to elucidate the confusing situation in the light WIMP region.
  
\section{Acknowledgements}
This work was supported in part by DOE grant DE-FG03-91ER40662 Task C.


\begin{thebibliography}{99}


\bibitem{Komatsu:2010fb}
  E.~Komatsu {\it et al.}  [WMAP Coll.],
 {\it  Astrophys.\ J.\ Suppl.}  {\bf 192} (2011) 18.
  
\bibitem{hannestad} 
  M. Kawasaki, K. Kohri, and N. Sugiyama, {\it Phys. Rev. Lett.} {\bf 82} (1999) 4168; {\it Phys. Rev.} {\bf D62} (2000)  023506; S.~Hannestad, {\it Phys. Rev.} {\bf D70}  (2004) 043506.

\bibitem{gg} 
  G.~Gelmini and P.~Gondolo,
  arXiv:1009.3690 [astro-ph.CO] and 
 {\it  Phys.\ Rev.}  {\bf D74} (2006)  023510;
G. Gelmini {\it et al.}
  {\it Phys. Rev.} {\bf D74} (2006)  083514;
  G. Gelmini,
 {\it Nucl.\ Phys.\ Proc.\ Suppl.} {\bf 194}, 63 (2009).

 \bibitem{Vogelsberger}
  M.~Vogelsberger {\it et.al.},
  arXiv:0812.0362[astro-ph];
  M.~Kamionkowski, S.~Kou- shiappas and M.~Kuhlen,
 {\it Phys.\ Rev.} {\bf D81} (2010) 043532;
  M.~Vogelsberger and S.~White,
  arXiv:1002.3162 [astro-ph.CO].

 \bibitem{Read}
  J. Read {\it et al.}
  arXiv:0803.2714[astro-ph];
  J. Read {\it et al.}
 {\it PoS} {\bf IDM2008} (2008) 048
  [arXiv:0901.2938[astro-ph.GA]];
  J. Read {\it et al.}
  arXiv:0902.0009[astro-ph].
  
\bibitem{Bernabei:2003za}
  R.~Bernabei {\it et al.}[DAMA Coll.],
 {\it Riv.\ Nuovo Cim.\ } {\bf 26N1}  (2003) 1.
  
\bibitem{Bernabei:2008yi}
  R.~Bernabei {\it et al.}  [DAMA Coll.],
  {\it Eur.\ Phys.\ J.} {\bf C56} (2008)  333.

\bibitem{Bernabei:2010mq}
  R.~Bernabei {\it et al.},
 {\it Eur.\ Phys.\ J.} {\bf C67} (2010) 39.
  
  \bibitem{Gelmini:2004gm}
  P. Gondolo and G. Gelmini
  {\it Phys. Rev.} {\bf D71}  (2005) 123520; also
  G. Gelmini and P. Gondolo
  hep-ph/0405278;
  G. Gelmini,
{\it J.Phys.Conf.Ser.} {\bf 39} (2006) 166.

  \bibitem{bottino} A. Bottino, N. Fornengo  and S. Scopel  
{\it Phys.~Rev.}~{\bf D67} (2003) 063519; A. Bottino~{\it et al.}
{\it Phys. Rev.} {\bf D69} (2004) 037302.

\bibitem{Fitzpatrick:2010em}
  A.~Fitzpatrick, D.~Hooper and K.~Zurek,
  {\it Phys.\ Rev.} {\bf D81}  (2010) 115005.

\bibitem{models}
  A.~Belikov~{\it et~al.}
  1009.0549 [hep-ph];
  J. Gunion, A. Belikov and D. Hooper,
  1009.2555 [hep-ph];
  M. Buckley, D. Hooper and T. Tait,
  1011.1499 [hep-ph].
  
\bibitem{CDMS-06}
D.S. Akerib {\it et al.} (CDMS coll.)  {\it Phys.\ Rev.\ Lett.} {\bf 96} (2006) 011302. 

\bibitem{Petriello:2008jj}
  F.~Petriello and K.~M.~Zurek,
  {\it JHEP} {\bf 0809} (2008) 057.
  
\bibitem{Bernabei:2007hw}
  R.~Bernabei {\it et al.},
  {\it Eur.\ Phys.\ J.}  {\bf C53} (2008) 205.

  \bibitem{Savage:2008er} 
  C.~Savage {\it et al.},
 {\it JCAP} {\bf 0904} (2009) 010.
  
   
\bibitem{Gemmell:1974ub}
  D. Gemmell,
 {\it Rev.\ Mod.\ Phys.}{\bf 46} (1974) 129;
D. Morgan and D. Van Vliet, 
 {\it Radiat. Effects and Defects in Solids} {\bf 8} (1971) 51;

 \bibitem{Hobler}
G. Hobler,
  {\it Radiat. Effects and Defects in Solids} {\bf 139} (1996) 21; 
G. Hobler,
  {\it NIM} {\bf B115} (1996) 323.

    \bibitem{altman} R. Altman {\it et al.}
 {\it Phys.\ Rev.} {\bf B7} (1973) 1743.

    \bibitem{Drobyshevski:2007zj}
  E.~M.~Drobyshevski,
{\it Mod.\ Phys.\ Lett.}  {\bf A23} (2008) 3077.
  
      \bibitem{BGG} N.~Bozorgnia, G.~Gelmini and P.~Gondolo,
{\it JCAP} {\bf 1011}  (2010) 019; see also
  {\it JCAP} {\bf 1011},  (2010) 028; 
   {\it JCAP} {\bf 1011}  (2010) 029; 
   [arXiv:1011.6006 [astro-ph.CO]];
  arXiv:1101.2876 [astro-ph.CO].
  
\bibitem{Savage:2010tg}
  C.~Savage {\it et al.}
  {\it Phys.\ Rev.} {\bf D83} (2011) 055002.
  
\bibitem{Ahmed:2009zw}
  Z.~Ahmed {\it et al.}  [The CDMS-II Coll.],
  {\it Science} {\bf 327} (2010) 1619.

\bibitem{Aalseth:2010vx}
  C.~E.~Aalseth {\it et al.}  [CoGeNT Coll.],
  {\it Phys.\ Rev.\ Lett.}  {\bf 106} (2011) 131301.
 
\bibitem{Chang:2010yk}
  S.~Chang {\it et al.}
  {\it JCAP}{\bf 1008}  (2010) 018.

\bibitem{Hooper:2010uy}
  D.~Hooper {\it et al.}
  {\it Phys.\ Rev.} {\bf D82} (2010) 123509.

  \bibitem{Manzur:2009hp}
  A.~Manzur {\it et al.},
  {\bf Phys.\ Rev.} {\bf C81}  (2010) 025808.
 
\bibitem{Seidel} W. Seidel, Talk at {\it WONDER}, LNGS, Italy, March 22-23 2010; also at {\it IDM 2010},
Montpellier, France, July 26-30, 2010.


\bibitem{Schwetz:2010gv}
  T.~Schwetz,
  arXiv:1011.5432 [hep-ph].
  
\bibitem{Akerib:2010pv}
  D.~Akerib {\it et al.} [CDMS Coll.],
  {\it Phys. Rev.} {\bf D82} (2010) 122004.
  
\bibitem{Ahmed:2010wy}
  Z.~Ahmed {\it et al.}  [CDMS-II Coll.],
  {\it Phys.\ Rev.\ Lett.} {\bf 106} (2011) 131302.
  
\bibitem{Aprile:2011hi}
  E.~Aprile {\it et al.}  [XENON100 Coll.],
  arXiv:1104.2549 [astro-ph.CO].  
  
\bibitem{Plante:2011hw}
  G.~Plante {\it et al.},
  arXiv:1104.2587 [nucl-ex].
  
  
\bibitem{Angle:2011th}
  J.~Angle {\it et al.}  [XENON10 Coll.],
  arXiv:1104.3088 [astro-ph.CO].
  
\bibitem{Aalseth:2011wp}
  C.~E.~Aalseth {\it et al.},
  arXiv:1106.0650 [astro-ph.CO].
  
\bibitem{Collar:2011wq}
  J.~I.~Collar,
  arXiv:1106.0653 [astro-ph.CO].
  
\bibitem{Hooper:2011hd}
  D.~Hooper and C.~Kelso,
  arXiv:1106.1066 [hep-ph].

\end{thebibliography}
\end{document}